\documentclass[aps,12pt,showpacs]{revtex4}
\usepackage[dvips]{graphics}
\usepackage[dvips]{graphicx}

\begin{document}
\title{Structure of the Isovector Dipole Resonance in Neutron-Rich $^{60}Ca$ Nuclei
and Direct Decay from Pygmy Resonance.}
\author{T.N. Leite and N. Teruya}
\affiliation{Departamento de F\'{\i}sica, Universidade Federal da Para\'{\i}ba \\
CP 5008, 58051-970 Jo\~{a}o Pessoa, Pb, Brazil}
\date{\today }

\begin{abstract}
The structure of the isovector dipole resonance in neutron-rich calcium
isotope, $^{60}Ca$, has been investigated by implementing a careful
treatment of the differences of neutron and proton radii in the continuum
random phase approximation ($RPA$). The calculations have taken into account
the current estimates of the neutron skin. The estimates of the escape
widths for direct neutron decay from the pygmy dipole resonance ($PDR$) were
shown rather wide, implicating a strong coupling to the continuum. The width
of the giant dipole resonance ($GDR$) was evaluated, bringing on a detailed
discussion about its microscopic structure.
\end{abstract}
\maketitle
\section{Introduction}

The investigation of the nuclei lying far from the $\beta ^{-}$ stability
line has been an interesting and active field of the nuclear physics in the
last two decades. In that region, a great number of exotic features are
observed like halo/skin formation, intruders levels, new magic numbers and
new kinds of collective excitation, the so-called soft and pygmy resonances.
These observations have forced the review of successful theoretical tools in
nuclei around the $\beta ^{-}$ stability valley \cite
{Tanihata,Caurier,Cot,Hart}.

The description of the microscopic structure of the exotic nuclei is a
current topic of study and several recent works had the concern of
describing the giant resonance ($GR$) in neutron-rich nuclei. The main
questions treated are the giant dipole resonance ($GDR$) behavior and the
appearance of the pygmy dipole resonance ($PDR$) in a nucleus with large
ratio of neutron to proton number ($N/Z$). The $PDR$ appears in medium and
heavy neutron-rich nuclei, and, within the hydrodynamic sense, they are
probably due to oscillation of the neutron excess against the core ($N=Z$
nucleus). The effects of the neutron excess in dipole resonances, in
neutron-rich nuclei, have been extensively studied in the literature by many groups 
\cite{Cham,Cata,Lanza,Reinhard,HSZ-ca1,Vretenar,Dang,Goriely1,Goriely2}, but
a complete understanding about the resonance structure has not been reached
up to now. Some questions about the nature of $GDR$ and $PDR$ are still kept
open. The preservation of the structure of the $GDR$ as the neutron number
increases, the composition of its decay modes and the need to go beyond $%
1p-1h$ configuration to explain it \cite{Dang} are some of these subjects of
theoretical interest. In addition to these interesting problems on nuclear
structure, the $PDR$ in neutron-rich nuclei plays an important subject in
the r-process nucleosynthesis \cite{Goriely1,Goriely2}. In this topic, the
dipole strength and the widths of the $PDR$ and $GDR$ are important
ingredients to understand the radiative neutron capture, because they are
directly related with the competition between the direct and statistical
mechanisms in the neutron capture process. In order to perform the studies
on neutron-rich nuclei, we choose to analyze the isovector dipole resonance
in the neutron-rich calcium isotope $^{60}Ca$. This nucleus was not reached
experimentally yet, but its microscopic structure has been investigated
recently in different versions of $RPA$ calculations \cite
{Cata,Lanza,Reinhard,HSZ-ca1}, relativistic $RPA$ ($RRPA$) \cite{Vretenar}
and phonon damping model ($PDM$) \cite{Dang}.

In this work we are interested in studying the role of continuum, as well as
the decay width, in the pygmy and giant dipole resonances in $^{60}Ca$
nucleus. In these extremely fragile nuclear systems, close to the neutron
drip line, the neutron density extends far away from the proton density,
forming the `neutron skin'. Thus, to perform these continuum calculations, we
have implemented a careful treatment of the differences of neutron and
proton radii in a version of the continuum $RPA$ approach of the Ref. \cite
{Teruya-npa}.

In the Sect. II of this paper, we describe the theoretical approach used in
the calculations. In the Sect. III we present and comment our results.

\section{Theoretical approach\qquad}

The continuum effects in our microscopic calculations are taken into account through
a discrete particle-hole basis which accomodates the single-particle resonance widths,
resulting in a diagonalization of $RPA$-like complex matrixes of standard size \cite{Teruya-npa}. The excited states are implemented in the particle-hole excitation space, and the open channels correspond to unbound particle-hole states with complex
energies, of which the imaginary parts give the single-particle escape
widths. We assume the nuclear hamiltonian of the form $H=H_{0}+\frac{1}{2}%
\sum_{i\neq j}V_{ij}$ with a mean field part $H_{0}$ and a residual two-body
force so that $(E_{\nu }-H)\mid \nu \rangle =0$ is satisfied for the nuclear
excited states $\mid \nu \rangle $ with energy $E_{\nu }$. Using the
orthogonal and complementary projectors $Q$ and $P$, the particle-hole space
can be splited onto normalized bound (or unbound) particle-hole states and
residual continuum states, respectively. Then, using the projection operator
formalism, we have the equations:

\begin{equation}
\text{ }[E_{\nu }-H_{QQ}-H_{QP}\frac{1}{E_{\nu }+i\eta -H_{PP}}H_{PQ}]Q\mid
\nu \rangle =H_{QP}\mid \chi ^{\dagger }\rangle \text{ ,}  \label{eq1}
\end{equation}
where $|\chi ^{\dagger }\rangle $ is the appropriate scattering solution of $%
H_{PP}$:

\begin{equation}
\text{ }[E_{\nu }-H_{PP}]\mid \chi ^{\dagger }\rangle =0\text{.}  \label{eq2}
\end{equation}
The relevant continuum effects in $P$ space is taken by accounting the
single-particle resonance in unbound particle-hole states (taken out for the 
$Q$ space) that is reached making the approach of ignoring the two body
force in $P$ space. In this way, the Eq.(\ref{eq2}) is a single-particle
equation and the continuum self-energy $(H_{0QP}\frac{1}{E_{\nu }+i\eta
-H_{0PP}}H_{0PQ})$ dresses the single-particle resonance with escape effects 
\cite{Teruya-prc,tnl}. Thus, we define the complex particle-hole modes $\mid
R_{n}\rangle $ with complex energies $\widehat{\varepsilon }_{n}=\varepsilon
_{n}-\frac{1}{2}\,i\,\Gamma _{n}^{\uparrow }$ which account for continuum
escape effects:

\begin{equation}
\lbrack \widehat{\varepsilon }_{n}-\hat{H}_{QQ}]\mid R_{n}\rangle =0\text{
and }[\widehat{\varepsilon }_{n}^{*}-\hat{H}_{QQ}^{\dagger }]\mid \widetilde{%
R}_{n}\rangle =0\text{ ,}  \label{eq3}
\end{equation}

\begin{equation}
\text{ }Q\mid \nu \rangle =\sum_{n}\frac{\mid R_{n}\rangle \langle 
\widetilde{R}_{n}\mid H_{QP}\mid \chi ^{\dagger }\rangle }{E_{\nu }-\widehat{%
\varepsilon }_{n}}\text{ ,}  \label{eq5}
\end{equation}
where $\hat{H}_{QQ}=H_{QQ}+H_{QP}\frac{1}{E_{\nu }+i\eta -H_{PP}}H_{PQ}$ and
the complex modes satisfy the orthonorma\-li\-zation relation: $\langle 
\widetilde{R}_{n^{^{\prime }}}\mid R_{n}\rangle =\delta _{nn^{^{\prime }}}$.
The matrix element in Eq.(\ref{eq5}) is an escape amplitude related to the
imaginary part of $\widehat{\varepsilon }_{n}$. The strength function $%
S_{F}(E)$\ $=\sum_{n}\frac{\Gamma _{n}^{\uparrow }}{2\pi }\frac{\left|
\left\langle \widetilde{R}_{n}\left| \widehat{F}_{\lambda }\right|
0\right\rangle \right| ^{2}}{(E-\varepsilon _{n})^{2}+(\Gamma _{n}^{\uparrow
}/2)^{2}}$ is calculated by making the approach that the excited states $\mid
\nu \rangle $\ can be well described by the $Q\mid \nu \rangle$ component.
In the particle-hole matrix element calculation of $1$-body operator $%
\widehat{F}_{\lambda }$ of the strength function $S_{F}(E)$, we have assumed
the form 
\begin{equation}
\widehat{F}_{JM}=e_{k}r^{J}Y_{JM}\text{ }  \label{eq-fjm-e1}
\end{equation}
where $e_{k}$ is the nucleon effective charge. For isovector dipole
transition, we have $e_{\nu (\pi )}=-\frac{eZ}{A}$\ $(\frac{eN}{A})$.

The complex particle-hole modes $\mid R_{n}\rangle $ in Eq.(\ref{eq3}) are
solved by a diagonalization of the discrete $RPA$ equations in $Q$ space:

\begin{equation}
\left( 
\begin{array}{cc}
A & B \\ 
-B & -A
\end{array}
\right) \left( 
\begin{array}{c}
X^{n} \\ 
Y^{n}
\end{array}
\right) =\widehat{\varepsilon }_{n}\left( 
\begin{array}{c}
X^{n} \\ 
Y^{n}
\end{array}
\right) \   \label{eq-rpa}
\end{equation}
where 
\begin{equation}
A_{php^{\prime }h^{\prime }}=(\widehat{\varepsilon }_{p}-\varepsilon
_{h})\delta _{pp^{\prime }}\delta _{hh^{\prime }}+V_{ph^{\prime }hp^{\prime
}}\text{ ; }B_{php^{\prime }h^{\prime }}=V_{pp^{\prime }hh^{\prime }}
\label{eq-aph-def}
\end{equation}
and $\widehat{\varepsilon }_{p}$ $(\widehat{\varepsilon }_{p}=\varepsilon
_{p}-\frac{1}{2}\,i\,\Gamma _{p})$ are the complex energies of the
single-particle resonances \cite{Teruya-prc,tnl}. Diagonalizing the complex
Eq.(\ref{eq-rpa}), we have the complex eigenvectors $\mid R_{n}\rangle $
(given by complex $X_{ph}^{n}$ and $Y_{ph}^{n}$ amplitudes) and complex
eigenvalues $\widehat{\varepsilon }_{n}$. Thus, in this discrete
particle-hole subspace, the escape width is associated with the contribution
of all allowed unbound particles coupled to their respective single-holes.
The partial escape width for each single-hole can be approximated for $\Gamma
_{h}^{n\uparrow }\simeq \sum_{p}\left| X_{ph}^{n}\right| ^{2}$\ $\Gamma _{p}$
, what gives a good estimate of the escape width composition for the
population of several single-holes in the residual nucleus.

The discrete single-particle energies are evaluated by solving the
Schr\"{o}dinger equation with Woods-Saxon potential, including the
centrifugal and Coulomb (as a uniformly charged sphere) terms. The positive
single-particle energy and its respective width are calculated in a
projection technique to continuum discretization approach \cite
{Teruya-prc,tnl}. The potential parameters of $^{60}Ca$ were adjusted
following the systematics of nucleon binding energy for this nucleus,
because there is no experimental data available for this nucleus. The $%
^{60}Ca$ is expected to have a small, but positive, neutron energy
separation and a large proton energy separation ($S_{n}\approx 3.5$ $MeV$
and $S_{p}\approx 25$ $MeV$). As a consequence, the single-particle potential of proton
should be deeper than the neutron one. Moreover, Since 60Ca is
already close to the neutron drip line, the next nucleus with full neutron
subshell, $^{70}Ca$, should be not stable against neutron emission. In
TABLE \ref{tab-vsp}, the potential parameters are displayed. The
single-particle energies calculated with these parameters are very similar
with those resulting of self-consistent Hartree-Fock ($HF$) calculation with 
$SIII$ Skyrme interaction \cite{Soojae,HSZ-60ca3}, except for the first
neutron excited state, namely $1g_{9/2}$ level, which was evaluated to be
unbound $(\varepsilon _{1g_{9/2}}\simeq 0.9$ $MeV)$, although, in other
calculations, it has been foreseen in a negative energy close to the border
of the potential well. However, these differences do not affect our results
for the calculated escape widths, because this is a very narrow resonance ($%
\Gamma _{1g_{9/2}}<1.0$ $keV$). 
%%%%%%%%%%%%%%%%%%%%%%%%%%%%%%%%%%%T1%%%%%%%%%%%%%%%%%%%%%%%%%%%%%%%%%%%
\begin{table}[tbp]
\caption{The Woods-Saxon central and spin-orbit parameters used in the
calculations. }
\label{tab-vsp}\centering
\begin{tabular}{ccc}
\hline\hline
\multicolumn{3}{c}{$V_{R}(r)=V_{0R}f(r)$ ;{\normalsize \ }$V_{ls}(r)=V_{0l%
\mathbf{s}}\left( \frac{h\hskip-.2em\llap{\protect\rule[1.1ex]{.325em}{.1ex}}%
\hskip.2em}{m_{\pi }c}\right) ^{2}\frac{1}{r}f^{\prime }(r) l \cdot s$} \\ 
\multicolumn{3}{c}{$f(r)=\frac{1}{1+e^{\left( r-R\right) /a}}${\normalsize \ 
}; $a=0.60\ fm$ ; $V_{0l\mathbf{s}}=6.54\ MeV$} \\ 
$Particle$ & $V_{0R}(MeV)$ & $R(fm)$ \\ \hline
$\nu $ & $54.5$ & $4.30$ \\ 
$\pi $ & $65.5$ & $4.62$ \\ \hline\hline
\end{tabular}
\end{table}
%%%%%%%%%%%%%%%%%%%%%%%%%%%%%%%%%%%%%%%%%%%%%%%%%%%%%%%%%%%%%%%%%%%%%%%%

The $RPA$ calculation is done by utilizing the Landau-Migdal residual
interaction: 
\begin{eqnarray}
V_{ph}(\overrightarrow{r_{1}},\overrightarrow{r_{2}})=C_{0}[f(r_{1})+f^{%
\prime }(r_{1})\overrightarrow{\tau _{1}}\cdot \overrightarrow{\tau _{2}}+ 
\nonumber \\
+\overrightarrow{\sigma _{1}}\cdot \overrightarrow{\sigma _{2}}\left(
g(r_{1})+g^{\prime }(r_{1})\overrightarrow{\tau _{1}}\cdot \overrightarrow{%
\tau _{2}}\right) ]\delta (\overrightarrow{r_{1}}-\overrightarrow{r_{2}})
\label{eq-lm-vres}
\end{eqnarray}
where $f$, $f^{\prime }$, $g$ and $g^{\prime }$ are dimensionless and
density dependent parameters: 
\begin{equation}
F(r)=F^{ex}+(F^{in}-F^{ex})\xi (r)\text{ .}  \label{eq-fsrho}
\end{equation}
The set of the interaction parameters is similar to that used in Ref. \cite
{KSTW,KST}, and it was adjusted to eliminate the spurious state $1^{-}$ (see
TABLE \ref{tab-para-vres}) and to reproduce the first $3^{-}$ excited state
around $2.0$ $MeV$.

We have considered a two-parameter Fermi distribution to represent the
density dependence $\xi (r)$ in Eq.(\ref{eq-fsrho}). In previous
calculations, we had assumed: 
\begin{equation}
\xi (r)=\frac{1}{1+e^{(r-R)/a}}\text{ ,}  \label{eq-f2p}
\end{equation}
where $R$ and $a$ are half-density radius and diffuseness, respectively.
Nevertheless, in the present context, it is more appropriate to separate this
function into neutron and proton parts: 
\begin{equation}
\xi (r)=\frac{N}{A}\xi _{\nu }(r)+\frac{Z}{A}\xi _{\pi }(r)\,\text{\ ,}
\label{eq-f2p-np}
\end{equation}
where each part is given by: 
\begin{equation}
\xi _{k}(r)=\frac{1}{1+e^{(r-R_{k})/a_{k}}}\text{ .}  \label{eq-f2p-k}
\end{equation}
Here $k=\nu (\pi )$ for neutron (proton). The Fermi distribution parameters
are adjusted to reproduce the neutron and proton expected root mean square
radii for the nucleus considered. It has been observed that the neutron root
mean square radius ($\sqrt{\langle r^{2}\rangle _{\nu }}$) becomes larger
than the respective proton radius ($\sqrt{\langle r^{2}\rangle _{\pi }}$) as
the neutron number increases, keeping constant the proton number.
The difference between those radii, $\Delta r_{\nu \pi }=\sqrt{\langle
r^{2}\rangle _{\nu }}-\sqrt{\langle r^{2}\rangle _{\pi }}$, gives the
measurement of the `neutron skin'. Some estimates of $\Delta r_{\nu \pi }$
have been extracted from experimental charged radii \cite{Trzc,Duf,Wohl,Mcca}. We
perform the $R_{k}$ and $a_{k}$ adjustment using the fact
that for a two-parameter Fermi distribution ($\rho _{k}(r)\propto \xi _{k}(r)
$) the mean square radius is given by \cite{Salcedo}: 
\begin{equation}
\langle r^{2}\rangle _{k}=\frac{\int_{0}^{\infty }\rho _{k}(r)r^{4}dr}{%
\int_{0}^{\infty }\rho _{k}(r)r^{2}dr}\approx \frac{3}{5}R_{k}^{2}+\frac{7}{5%
}\pi ^{2}a_{k}^{2}\,\text{\ .}  \label{eq-rmsk}
\end{equation}

%%%%%%%%%%%%%%%%%%%%%%%%%%%%%%%%%%%T2%%%%%%%%%%%%%%%%%%%%%%%%%%%%%%%%%%%
\begin{table}[tbp]
\caption{Parameters of harmonic oscillator radial wavefunctions and residual
interaction. }
\label{tab-para-vres}\centering
\begin{tabular}{ccccccc}
\hline\hline
\multicolumn{7}{c}{$^{60}Ca$} \\ 
$\sqrt{\langle r^{2}\rangle _{\nu }}(fm)$ & $R_{\nu }(fm)$ & $a_{\nu }(fm)$
& $\sqrt{\langle r^{2}\rangle _{\pi }}(fm)$ & $R_{\pi }(fm)$ & $a_{\pi }(fm)$
& $\sqrt{\langle r^{2}\rangle }(fm)$ \\ 
$4.03$ & $4.34$ & $0.60$ & $3.62$ & $3.68$ & $0.60$ & $3.90$ \\ \hline
\multicolumn{3}{c}{$b_{\nu }(fm)$} & \multicolumn{4}{c}{$b_{\pi }(fm)$} \\ 
\multicolumn{3}{c}{$2.09$} & \multicolumn{4}{c}{$2.11$} \\ \hline
$C_{0}(MeV$ $fm^{3})$ & $f^{in}$ & $f^{ex}$ & $f^{\prime in}$ & $f^{\prime
ex}$ & $g$ & $g^{\prime }$ \\ 
$300.0$ & $-0.002$ & $-2.1$ & $0.76$ & $2.3$ & $0.51$ & $0.70$ \\ 
\hline\hline
\end{tabular}
\end{table}
%%%%%%%%%%%%%%%%%%%%%%%%%%%%%%%%%%%%%%%%%%%%%%%%%%%%%%%%%%%%%%%%%%%%%%%%
In our $RPA$ calculation, we represent the radial single-particle orbits by
harmonic oscillator radial wave functions $R_{i}(r)$ which are characterized
by a size parameter $b$, and we have adopted a two-parameters Fermi density
dependence in the residual interaction. Since the neutron and proton
densities are too different in exotic nuclei, we must pay special
attention with the parameter $b$. Therefore, we have performed a careful choice of the parameter $b$ to analyze the microscopic structure of this exotic nucleus. In this way, using
the radial distribution of harmonic oscillator, $\rho _{k}(r)=\rho
_{k}^{0}\sum_{\varepsilon _{i}\leq \varepsilon _{F}}\left( 2j_{i}+1\right)
R_{i}^{2}(r;b_{k})$, in order to calculate $\langle r^{2}\rangle _{k}$, we
satisfy this requisition by assuming different size parameters for neutrons
and protons, 
\begin{equation}
b_{k}^{2}\approx \frac{4}{(3)^{\frac{4}{3}}}\langle r^{2}\rangle
_{k}(X_{k})^{-\frac{1}{3}}\text{ ,}  \label{eq-bk-rmsk}
\end{equation}
where $X_{k}=N$\ $(Z)$\ for $k=\nu (\pi )$\ and $\langle r^{2}\rangle _{k}$
is given by Eq.(\ref{eq-rmsk}). In the TABLE \ref{tab-para-vres}, we have
displayed the $R_{k}$ and $a_{k}$ adjustment based on some recent radii estimates 
\cite{ZMTTZ,Meng,Suga}.

\section{Discussion and Conclusions}

\begin{figure}[tbp]
\begin{center}
\includegraphics[angle=270,width=\textwidth]{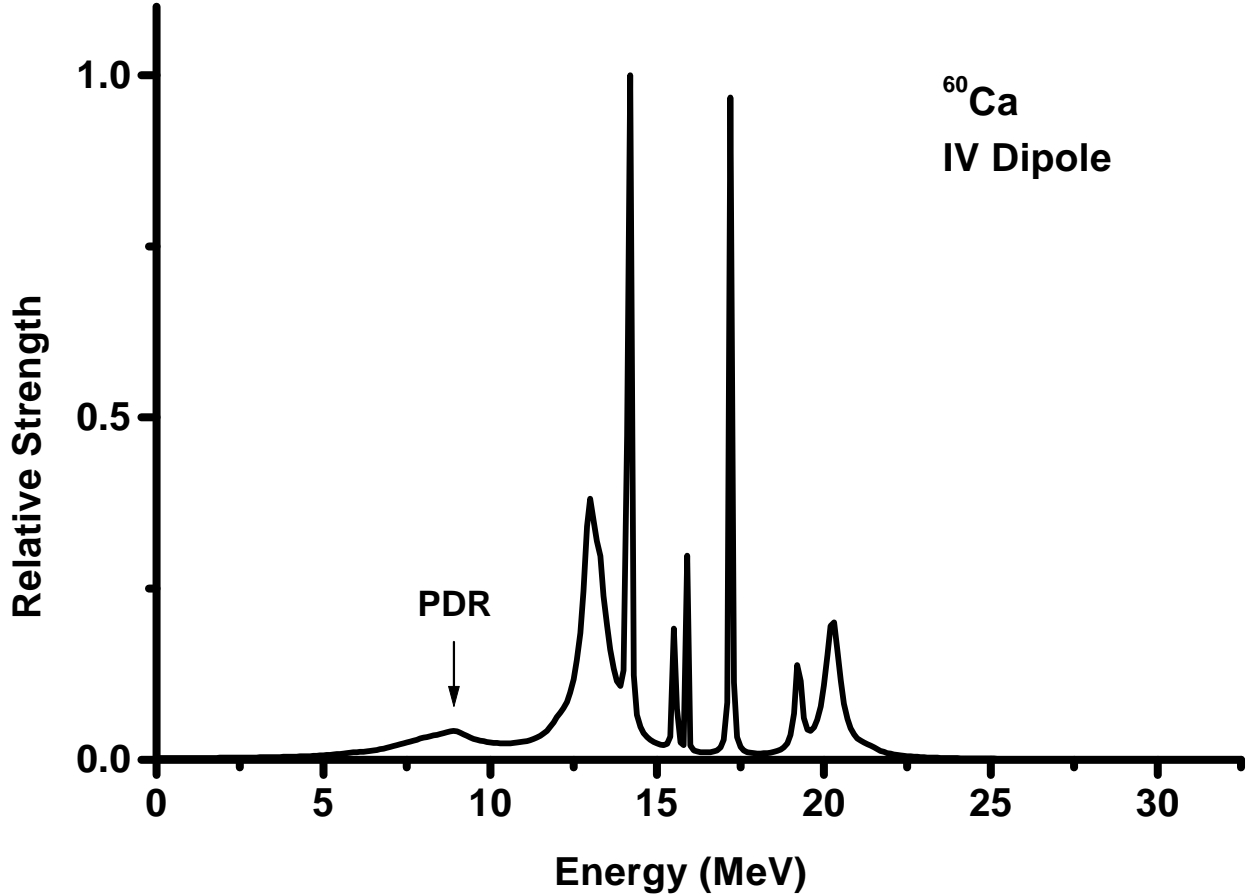}
\end{center}
\caption{The calculated strength function $S_{F}(E)$ by continuum $RPA$
approach for the $E1$ excitation in $^{60}Ca$ nucleus. The wide peak in $%
\sim 8.6$ $MeV$ is in the energy interval where the $PDR$ is expected.}
\label{fig-sf}
\end{figure}

In this section, we present and discuss some results which we have obtained
by using $1p-1h$ continuum $RPA$ approach, as described in the previous section, for
isovector dipole electric transition for $^{60}Ca$. The
single-particle(hole) energy levels were taken from the $1s_{1/2}$ hole up
to $1g_{7/2}$ particle orbits. In this case, the core ($N=Z=20$) is formed by
neutrons and protons filling the energy levels up to $sd$ shell and the
neutrons excess occupying the $pf$ shell above the core. The calculated
strength function $S_{F}(E)$ is displayed in the FIG. \ref{fig-sf}. 
%%%%%%%%%%%%%%%%%%%%%%%%%%%%%%%%%%%F1%%%%%%%%%%%%%%%%%%%%%%%%%%%%%%%%%%%
%%%%%%%%%%%%%%%%%%%%%%%%%%%%%%%%%%%%%%%%%%%%%%%%%%%%%%%%%%%%%%%%%%%%%%%%

In the $^{40}Ca$ nucleus, the $GDR$ is located around an energy ($\sim 20\
MeV$ ) above the proton ($S_{p}\approx 8\ MeV$) and neutron separation
energy ($S_{n}\approx 16\ MeV$) \cite{Brajnik-40ca,Diesener-40ca}. Consequently, the
proton and neutron emission competes by decay of this mode. Taking into account
the neutron-rich nucleus ($^{60}Ca$), the theoretical predictions have
displayed that $PDR$ appears below the $GDR$ as a result of neutron excess.
The $PDR$ and $GDR$ are expected to appear below $10$ $MeV$ and $15-20$ $MeV$
\cite{Cata,Lanza,Reinhard,HSZ-ca1,Vretenar,Dang}, respectively. The $GDR$ is
shifted down due to the increase of the mass. However, its microscopic structure
is found to be composed mainly by the single-holes that belong to the most
internal structure of the core, showing that they belong to the same
category of excitations of the $^{40}Ca$. As we have considered $%
S_{n}\approx 3.5$ $MeV$ and $S_{p}\approx 25$ $MeV$ for the $^{60}Ca$
nucleus, the one-neutron channel is presumed to be open in the $PDR$ region.

%%%%%%%%%%%%%%%%%%%%%%%%%%%%%%%%%%%T3%%%%%%%%%%%%%%%%%%%%%%%%%%%%%%%%%%%
\begin{table}[tbp]
\caption{The evaluated escape widths, and the estimates of the partial
escape widths for each single-hole, of the three main peaks that compose the 
$PDR$ in $\sim 8.6$ $MeV$ of excitation energy (see FIG. \ref{fig-sf})
exhausting about 8$\%$ of the $EWSR$. }
\label{tab-escape}\centering
\begin{tabular}{lll}
\hline\hline
$\varepsilon _{n}-\frac{1}{2}\,i\,\Gamma _{n}^{\uparrow }(MeV)$ & $%
single-hole$ & $\Gamma _{h}^{n\uparrow }(MeV)$ \\ \hline
$8.05-i\,1.29$ & $(1f_{5/2})_{\nu }$ & $2.26$ \\ 
& $(2p_{1/2})_{\nu }$ & $0.39$ \\ 
& $(2p_{3/2})_{\nu }$ & $0.02$ \\ \hline
$8.69-i\,1.35$ & $(1f_{5/2})_{\nu }$ & $0.05$ \\ 
& $(2p_{1/2})_{\nu }$ & $2.65$ \\ 
& $(2p_{3/2})_{\nu }$ & $0.20$ \\ \hline
$8.94-i\,0.40$ & $(1f_{5/2})_{\nu }$ & $0.04$ \\ 
& $(2p_{1/2})_{\nu }$ & $0.09$ \\ 
& $(2p_{3/2})_{\nu }$ & $0.71$ \\ \hline\hline
\end{tabular}
\end{table}
%%%%%%%%%%%%%%%%%%%%%%%%%%%%%%%%%%%%%%%%%%%%%%%%%%%%%%%%%%%%%%%%%%%%%%%%%%%%%%%%%%%%%%%%%%%%%
According to previous $RPA$ calculations \cite{Cata,Lanza,Reinhard,HSZ-ca1}%
, our calculations also predict a considerable strength in the energy region
below $10$ $MeV$, what can be observed by the presence of the broad neutron
peak in low-lying energy in FIG. \ref{fig-sf}. The broad width is due to the
fact that the energy of the resonance to be above of the small neutron
separation energy, implicating a strong coupling of the external neutrons to
the continuum region. These low-lying energy states are the natural
candidates for $PDR$ because they have a dominant contribution of `neutron
skin' ($pf$ shell), in agreement with hydrodynamic interpretation. In FIG. 
\ref{fig-sf} the wide peak in $\sim 8.6$ $MeV$ is composed by the overlap of
three main peaks (see TABLE \ref{tab-escape}) exhausting about $8\%$ of the
Energy Weighted Sum Rule ($EWSR$), which are composed by transitions involving neutron of `skin' ($2p$ and $1f\rightarrow 3s$, $2d$ and $1g$). This aspect becomes more clear
when the structure of the $RPA$ wave function of these peaks is observed.
The wave functions of the states at $8.05$ $MeV$, $8.69$ $MeV$ and $8.94$ $MeV$
are composed by the main neutron $1p-1h$ configurations: $\left( 76\%\left|
1f_{5/2}^{-1}2d_{3/2}\right\rangle _{\nu }+14\%\left|
2p_{1/2}^{-1}3s_{1/2}\right\rangle _{\nu }\right) ${\sl , }$\left(
93\%\left| 2p_{1/2}^{-1}2d_{3/2}\right\rangle _{\nu }+6\%\left|
2p_{3/2}^{-1}2d_{3/2}\right\rangle _{\nu }\right) $ and $\left(
88\%\left| 2p_{3/2}^{-1}2d_{5/2}\right\rangle _{\nu }+9\%\left|
1f_{7/2}^{-1}1g_{9/2}\right\rangle _{\nu }\right) $, respectively, showing that each one of these peaks has a do\-mi\-nance of two $1p-1h$ configurations. These results are in agreement with the $HF+RRPA$ calculation of the Ref. \cite{Vretenar}, where they have noticed that the one, or at most two neutron $1p-1h$ configurations, determine the structure of the low-energy states for isovector dipole resonance in $^{60}Ca$, in
contrast to the structure of the collective states which are characterized
by a coherent superposition of many $1p-1h$ excitations.
However, the $PDR$ is composed by the overlap of those three main states, and
the collectivity can be understood through the participation of all
particle-hole components that belong to this overlap.
%%%%%%%%%%%%%%%%%%%%%%%%%%%%%%%%%%%T4%%%%%%%%%%%%%%%%%%%%%%%%%%%%%%%%%%%
\begin{table}[tbp]
\caption{Estimates for the mean values of the escape widths for the $PDR$. }
\label{tab-escapemedio}\centering
\begin{tabular}{lll}
\hline\hline
$\overline{\varepsilon }_{n}-\frac{1}{2}\,i\,\overline{\Gamma }%
_{n}^{\uparrow }(MeV)$ & $single-hole$ & $\overline{\Gamma }_{h}^{n\uparrow
}(MeV)$ \\ \hline
$8.6-i\,1.3$ & $(1f_{5/2})_{\nu }$ & $0.7$ \\ 
& $(2p_{1/2})_{\nu }$ & $1.6$ \\ 
& $(2p_{3/2})_{\nu }$ & $0.2$ \\ \hline\hline
\end{tabular}
\end{table}
%%%%%%%%%%%%%%%%%%%%%%%%%%%%%%%%%%%%%%%%%%%%%%%%%%%%%%%%%%%%%%%%%%%%%%%%%%%%%%%%%%%%%%%%%%%%%
The estimates of the partial widths $\Gamma _{h}^{n\uparrow }$
of the main neutrons single-holes that populate the $PDR$ are presented in TABLE \ref
{tab-escape}. We have also evaluated an average width for the $PDR$, calculated as the
weighted average of the widths of the peaks that compose it (see TABLE \ref
{tab-escapemedio}), the weights being the intensities of each peak relative
to the $8\%$ of the $EWSR$ that they exhaust.
In comparision to these predictions to the $PDR$, we can note an apparent
lack of neutron escape in the region of energy around the $15$ $MeV$%
, namely $GDR$ region, where the peaks are very narrow (see FIG. \ref{fig-sf}). These peaks are constituted principally by bound $1p-1h$ configurations of protons and their narrow widths are due to the small neutrons amplitudes in this energy interval, reflecting a
unexpected suppression of the neutron emission. This result is mainly due to
the change of status of the neutrons single-particle levels, because those
neutrons of $pf$ shell, that belonged to the configurations of the particle
states in $^{40}Ca$ nucleus, have become hole states in the neutron-rich
nucleus, and their contributions for the $GDR$ of the $^{60}Ca$ are small
(they are mainly responsible for the $PDR$ appearance). Furthermore, the
composition of these narrow peaks is owed mainly to the bound $1p-1h$ pairs
of protons. This neutron lack can indicate that the $GDR$ decay should also have more complex contributions than $1p-1h$ excitations.
%frase original:
%This neutron lack can indicate that the $GDR$ decay should also
%have another contribution that goes by more complex excitations than $1p-1h$.
%mudando para
%This neutron lack can indicate that the $GDR$ decay should also have more complex %contributions than $1p-1h$ excitations.
%ou
%This neutron lack can indicate that the $GDR$ decay should also have contributions beyond %$1p-1h$ excitations.
%%%%%%%%%%%%%%F2%%%%%%%%%%%%%%%%%%%%%%%%%%%%%%%%%%%
\begin{figure}[tbp]
\begin{center}
\includegraphics[width=\textwidth]{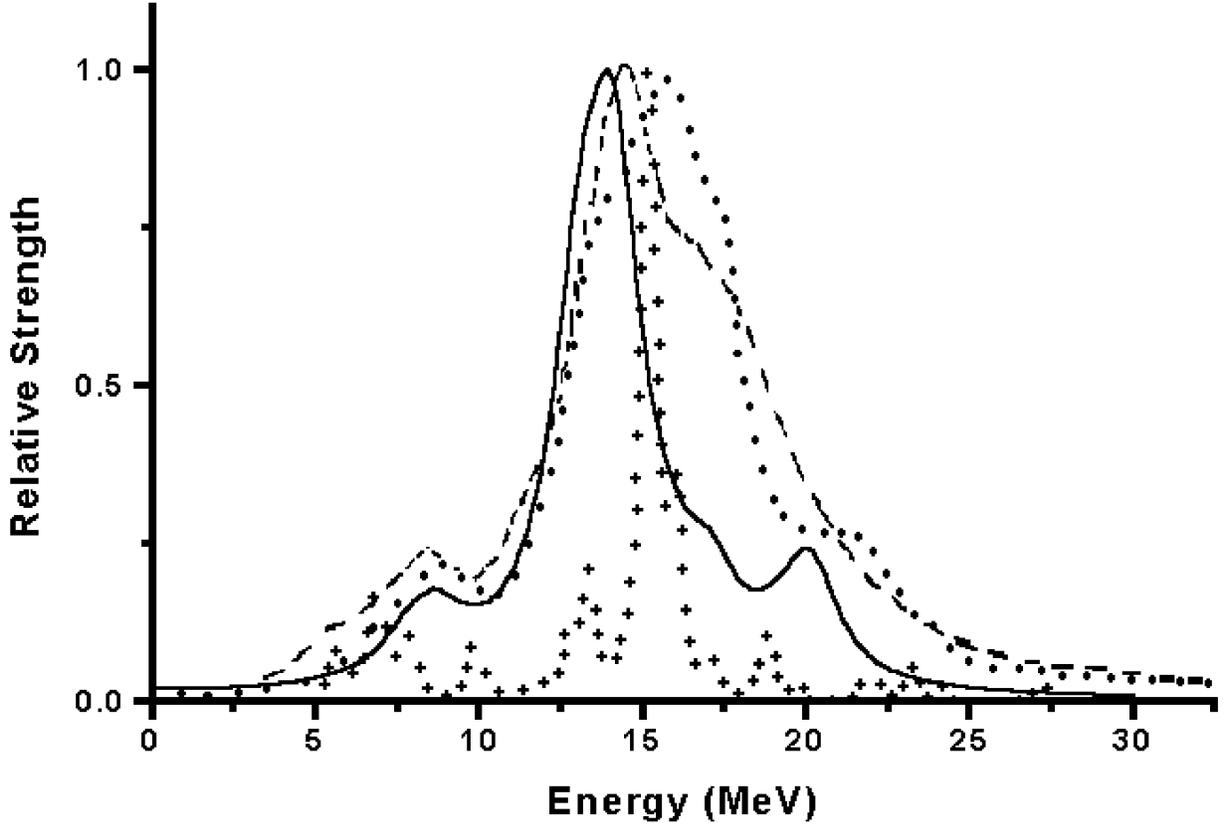}
\end{center}
\caption{Comparison among our results (solid line) with the one of other works: $%
HF+RPA$ calculations by Catara et. al. \protect\cite{Cata} (dotted line),
Hamamoto et. al \protect\cite{HSZ-ca1} (dashed line) and $HF+RRPA$ calculations
by Vretenar et. al. \protect\cite{Vretenar} (crosses). Our calculation is
performed by including an arbitrary constant width $\Gamma =1.0$ $MeV$,
instead of calculated single-particle widths used in the calculation
presented in FIG. \ref{fig-sf}.}
\label{fig-others}
\end{figure}
%%%%%%%%%%%%%%%%%%%%%%%%%%%%%%%%%%%%%%%%%%%%%%%%%%%%%%%%%%%%%%%%%%%%%%%%
The analysis of the underlying structure of the resonance in the exotic
nuclei may help us understand such disagreements because the $GDR$ could
have a more complex structure than $1p-1h$ configurations, what could
contribute to the spreading width. To simulate these more complex structures
of the $GDR$, we display in the FIG. \ref{fig-others} our calculation using a
constant width $\Gamma =1.0$\ $MeV$, instead of the single-particle widths
used in the calculation presented in the FIG. \ref{fig-sf}, and we compare
the results with the one of the Refs. \cite{Cata,HSZ-ca1,Vretenar}.
Moreover, the two-neutron separation energy ($S_{2n}\approx 7$\ $MeV$) is
expected to be too smaller than $S_{p}$ in the neutron rich nuclei \cite
{Meng}. Then the two-neutron channel is presumed to be open above the $PDR$
and below the $GDR$ energy, and it is expected to constitute an important
part of $GDR$ decay. As our continuum $RPA$ calculations have taken into
account only $1p-1h$ configurations, this two-neutron emission is not
considered. Therefore, the inclusion of $2p-2h$ configuration in the
calculations would enhance the neutron emission by the $GDR$.

At present, it may be difficult, or else impossible, to make the experimental
measures of these widths, but it could be worthwhile doing a theoretical
effort for a better understanding of the microscopic structure of these
exotic nuclei, that would allow us to accomplish some estimates. In Ref. 
\cite{Goriely2} the low energy dipole strength, in capture cross section
calculation, is folded by a Lorentzian curve of width $\Gamma _{PDR}=\Gamma
_{GDR}\left( E_{PDR}/E_{GDR}\right) ^{2}$, and it is added to the $GDR$
strength to satisfy the classical sum rule for the total $E1$ strength. We
could use this relation to perform an evaluation of the $\Gamma _{GDR}$. In
TABLE \ref{tab-escapemedio} we show the mean values of the energy ($8.6$ $MeV
$) and escape width ($2.6$ $MeV$) for the $PDR$. The large value of the
escape width could suggest that it is the main component of its total width,
or $\Gamma _{PDR}=2.6$ $MeV$. On the other hand, our calculation gives a
narrow $GDR$ centered in $15$ $MeV$. In spite of our calculation does not
contain 2p-2h or more complex excitations, we can use these results for $PDR$
and $GDR$, together the above expression \cite{Goriely2} for the $\Gamma
_{PDR}$, to evaluate the width $\Gamma _{GDR}$. Proceeding this way, we
obtain $\Gamma _{GDR}=7.9MeV$. This result shows that the $GDR$ is a very
wide resonance, but with a small $1p-1h$ escape width, and reinforces the above
statement about the possible complexity of the $GDR$ structure. The
connection of these two different calculations supplies a value for the $%
\Gamma _{GDR}$ that is close to the result obtained in the $PDM$
calculations of the Ref. \cite{Dang}.

In short, our approach has been describing consistently the $1p-1h$
microscopic nature of the $PDR$ excitation mode. The transitions in this
region are composed by $1p-1h$ excitations involving neutrons of `skin'. The
broad peaks are due to the coupling to continuum, which is extremely
important in these fragile systems. Furthermore, we have exhibited an
estimate about the direct escape of neutron from $PDR$. Regarding the $GDR$, due to small escape width obtained in our calculations, the results
indicate that the more complicated excitations than $1p-1h$ should also be
important for the description of its microscopic structure.

\section{Acknowledgments}

\qquad This work was supported in part by Conselho Nacional de Desenvolvimento
Cient\'{i}fico e Tecnol\'{o}gico (CNPq), Brazil.

\end{document}